\documentclass[12pt]{iopart}

\usepackage[dvips]{graphicx}

\newcommand{\ssnnm}{\sqrt{s_{NN}}}
\newcommand{\ssnn}{$\ssnnm$}
\newcommand{\snn}[1]{{\ssnn} = #1 GeV}
\newcommand{\gev}{{\rm \; GeV}}

\newcommand{\mea}[1]{\left< #1 \right>}
\newcommand{\pair}[1]{#1 \overline{#1}}
\newcommand{\dpair}[1]{$\overline{#1} / #1$}

\newcommand{\Xim}{$\Xi^-$}
\newcommand{\Xip}{$\overline{\Xi}^+$}
\newcommand{\pim}{$\pi^-$}
\newcommand{\pip}{$\pi^+$}
\newcommand{\tld}{$\sim$}
\newcommand{\muB}{\mu_{\rm B}}

\begin{document}

\title[STAR Strangeness Results from \ssnn = 130  and 200 GeV Au+Au Collisions]
{STAR Strangeness Results from \snn{130} Au+Au Collisions
(and first results from 200 GeV)}

\author{G Van Buren\dag~for the STAR Collaboration
\footnote{For full collaboration author list, see L. Barnby for the STAR Collaboration, these proceedings.}}
\address{\dag~Brookhaven National Laboratory, Upton, NY 11973-5000, USA}

\ead{gene@bnl.gov}

\begin{abstract}
The STAR Experiment at RHIC is capable of a wide variety of measurements
of the production of strange hadrons in nuclear collisions.
Measurements of the relative production of strange baryons, antibaryons,
and kaons can shed light on the baryon densities achieved in these
collisions and on the validity of models for production yields. We will present
here preliminary results on these measurements at RHIC energies of
\snn{130} and 200 GeV and
discuss comparisons to models.
\end{abstract}

\submitto{\JPG}

\section{Introduction}

Hadron production in a wide variety of colliding systems and energies has
been well-described by statistical models~\cite{stat}.  Deviations from these models
may be an indication of new physics, perhaps the production of quark gluon
plasma~\cite{qgp}.  It is therefore important to test for such deviations in new data
from RHIC.

The STAR experiment can measure a wide variety of hadron yields at mid-rapidity.
While statistical models for particle production generally ask for global yields
($4\pi$ measurements), this may not be necessary at RHIC.
Particle yields at high rapidities are
expected to have significant content from the initial colliding nuclei, leading to
truly different sources with different statistical descriptions at mid-rapidity and
at high rapidity.  Because beam rapidity is large (\tld5-6) in the center-of-mass frame
at RHIC energies, contamination from the high-rapidity hadron sources should
be small at mid-rapidity.  This should allow a rather clean measurement of the
source there, provided that global equilibrium is approximated by local equilibrium.

In order to allow the use of data from different RHIC experiments in model fits
and comparisons, taken at varying definitions of a central-collision selection, it
is helpful to take particle ratios instead of absolute yields. This has the side
benefit of also reducing some common-mode errors in particle yield measurements.

\section{Models \& \snn{130} Data}

\subsection{Thermal Statistical Models}
\label{se:stat}

For a thermal model to make predictions about particle ratios,
it must have a description of the equation of state (EOS) of the system
at chemical freezeout.
In the field of heavy ion physics, the commonly used parameters to describe the
EOS are those of baryo-chemical
potential ($\muB$) and temperature ($T$), along with the requirement that
net strangeness is zero.  The authors of
Ref.~\cite{PBM1} chose two phenomenological
parameterizations to provide these quantities for central
heavy ion collisions:
\begin{eqnarray}
\label{eq:mub_s}
\muB(\ssnnm) \simeq \frac{1.27 \gev}{1 + \ssnnm/(4.3 \gev)} \\
\label{eq:e_over_n}
\mea{E} /\mea{N} \simeq 1 \gev
\end{eqnarray}

Constrained by the strangeness neutrality,
$\mea{E} / \mea{N}$ depends only upon $\muB$ and $T$~\cite{stat}, so $T$ is
determined by the above equations for any collision energy \ssnn.
Via a canonical partition function (necessary, at least at low energies,
to ensure exact strangeness conservation~\cite{canon}), they arrive at ratios of particles as a
function of \ssnn.  Their predictions for the energy dependence of
these ratios hold rather well at AGS and SPS energies~\cite{red}, and lead
to an expected slight drop in ratios of strange to non-strange particles going from
SPS to RHIC energies (see Ref.~\cite{PBM1} for an explanation).
The drop is most notable for the ratio of $\Lambda$ to $\pi$ yields.
Although preliminary $\Lambda$ yields have been presented by STAR
at this conference~\cite{matt},
the proper comparison requires full understanding of feeddown contributions
to the $\Lambda$ yields and is still in progress.

Though not as large, the \Xim/\pip~ratio is still expected to drop by more than 30\% in this model.
The $\Xi$ does not have significant feeddown corrections, so the comparison is easier.
STAR has also presented preliminary \Xim~and \Xip~yields at this conference~\cite{jecc},
and using a STAR preliminary \pim~yield~\cite{manuel} (which we know approximates the
\pip~yield well~\cite{piratio}), we can evaluate the prediction.

The result is a measured ratio nearly twice as large as that predicted! One must, of course, consider
the significance of the difference. The statistical error on the ratio is \tld7\%, with a systematic
error expected to be \tld20\%, so the data point is close to two standard deviations away
from the prediction.

\begin{figure}
\begin{center}
\vspace{-0.2cm}
\includegraphics[height=5.0cm]{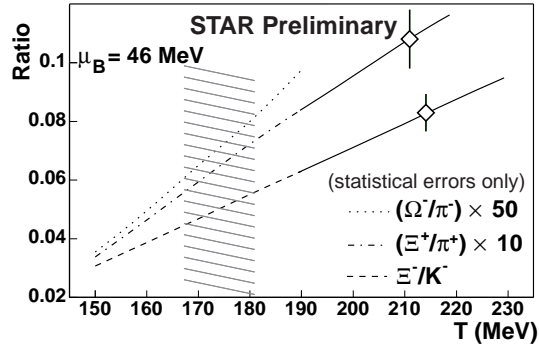}
\vspace{-0.7cm}
\end{center}
\caption{Sensitivity of the statistical model fit temperature (T) to
various particle ratios. Preliminary ratios measured by STAR are shown along
hand-drawn extrapolations (solid lines) of the theoretical curves (dashed lines). The hashed region
represents the temperature range of a fit to 130 GeV RHIC data~\cite{PBM2}.}
\label{fi:sensitivity}
\end{figure}

Conversely, one might ask how sensitive the statistical models are to
these ratios.  In Ref.~\cite{PBM2}, the authors apply a grand canonical model to many particle
ratios from 130 GeV RHIC data.  They find that the model fits the measured data well (with chemical
potential not too different from the parameterization of \eref{eq:mub_s}), and go on
to show the expected range of several strange/non-strange particle ratios given
their fit parameters, as well as the sensitivity of these ratios to their temperature parameter T.
This sensitivity is demonstrated by the theoretical curves shown in \fref{fi:sensitivity}.
The STAR preliminary values for some of these ratios are also shown.
It is worthwhile to note that the two measured ratios shown are both too high by approximately
the same amount from the model calculations simply because the $K/\pi$ ratios are
in the fit, leaving only the over-abundance of \Xim~and \Xip~to offset the ratios (the relative
abundance between \Xim~and \Xip~being rather well-constrained by the chemical potential).
It is clear that these models cannot accommodate the preliminary yields of the \Xim~and \Xip.

\subsection{Quark Coalescence Model}

ALCOR (a quark coalescence model) has demonstrated that it can fit STAR mid-rapidity
antiparticle/particle ratios involving strangeness well~\cite{qc1}. Because it is also
a statistical model, although not an equilibrium model, these ratios
are driven primarily by the relative difference in the abundance of quarks and antiquarks.
Transported numbers of valence quarks (or baryon number)
to mid-rapidity are smaller than those from pair production processes at RHIC
collision energies, so this relative difference becomes smaller, driving the ratios
towards one and diminishing statistical sensitivity to small differences.
It is therefore not surprising that all statistical models fit antiparticle/particle ratios
rather well at these (and higher) energies.

More powerful in distinguishing between models are ratios between highly
dissimilar particle species. Differences may include quark flavor content, mass, spin, and
general abundance, for example (although these are not completely independent quantities).
The $K/\pi$ ratios exemplify some of these dissimilarities, and already have shown
good agreement with ALCOR (given some assumptions on
$\pair{s}$ production)~\cite{qc1}. Even more highly dissimilar are the
$\Xi/\pi$ ratios. ALCOR over-predicts the ratio by more than 10\%,
although taking into account the \tld20\% systematic error of the measured ratio brings the data
into agreement with the model for the $\Xi$ yields~\cite{qc2}.

\section{First Results from \snn{200} Data}

\subsection{Quality of New Data}

\begin{figure}
\begin{center}
\vspace{-0.2cm}
\includegraphics[height=3.2cm]{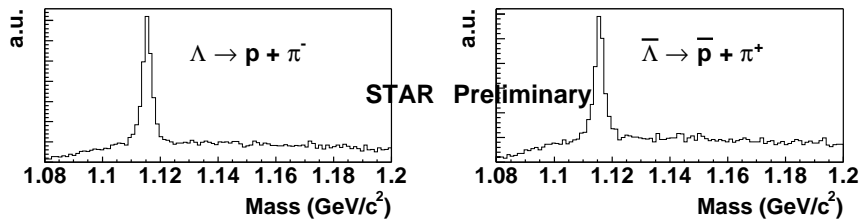}
\vspace{-0.6cm}
\end{center}
\caption{Preliminary invariant mass peaks for $\Lambda$ and $\bar{\Lambda}$ in minimum bias
\snn{200} Au+Au collisions.}
\label{fi:200gev}
\end{figure}

\Fref{fi:200gev} shows the preliminary invariant mass peaks for
$\Lambda$ and $\bar{\Lambda}$ in 50,000 minimum bias \snn{200} events
taken from the 2001 run. It is simple to extract a preliminary
\dpair{\Lambda} ratio from this
data because most of the inefficiencies cancel out, resulting in a
value only slightly higher than the ratio from
130 GeV~\cite{matt}.  However, it is
important to note that this number is not corrected for absorption of antiprotons
in the detector apparatus.  The correction for this data is now quite significant
as there is more material in the detector due to the addition of
the Silicon Vertex Tracker and associated components, raising the
ratio further.

\subsection{Comparison with Model Prediction}

Using the parameterization given in \eref{eq:mub_s}, $\muB$
at 200 GeV should be \tld26.7 MeV. The authors of Ref.~\cite{bec} use a
"strangeness-canonical" scheme (a grand canonical model modified to include
exact strangeness conservation) and the parameterization of
\eref{eq:e_over_n} to arrive at predictions for antiparticle/particle
ratios as a function of $\muB$. One can read off the
value from their calculations for a small range of temperatures that
\dpair{\Lambda} should be slightly above or close to 0.8.  It appears
that the preliminary STAR data from 200 GeV will be in this range after
correcting for absorption. This says little about the statistical model because
of its loss of sensitivity to antiparticle/particle ratios at these energies, but
it indicates that the parameterization of \eref{eq:mub_s} is at least close.
If this parameterization is correct, it implies that we must go to very high
collision energies to legitimately approximate a net-baryon-free region at mid-rapidity.

\section{Conclusions}

All statistical models do rather well at fitting antiparticle/particle ratios at RHIC. This
comes in part from the approach of these ratios to a value of one. Not only
do the models begin to lose sensitivity to these ratios as this happens,
the experimental measures of these ratios also lose their resolving
power due to inherent measurement errors. Ratios between highly dissimilar
particles become more and more important for distinguishing between
models, the $\Xi/\pi$ ratios being a good example.

It is clear that the preliminary measurements of the $\Xi$ yields from STAR in
central Au+Au collisions at \snn{130} are not in agreement with thermodynamic
statistical models. If the multiple strange quarks in the $\Xi$
are the source of this enhancement, it would be very interesting to see the
results of the $\Omega$ yields, even if the measurement errors are
very large.
Nonetheless, the quark coalescence model appears to
provide a better match for the values of $\Xi/\pi$, without sacrificing
other particle ratios. Because the prediction of this model is
in clear disagreement with the thermal models,
it will be very important to finalize these results.

Finally, the parameterization given by
\eref{eq:mub_s} and \eref{eq:e_over_n} appear to hold at both
RHIC energies run so far, indicating that these energies are still not
producing a net-baryon-free region at mid-rapidity in Au+Au collisions.

\end{document}